# Analysis and Design of Ultra Thin Electromagnetic Absorbers Comprising Resistively Loaded High Impedance Surfaces

Filippo Costa *Student Member, IEEE* , Agostino Monorchio, *Senior Member, IEEE*, and Giuliano Manara, *Fellow, IEEE*

*Abstract*—High-Impedance Surfaces (HIS) comprising lossy Frequency Selective Surfaces (FSS) are employed to design thin electromagnetic absorbers. The structure, despite its typical resonant behavior, is able to perform a very wideband absorption in a reduced thickness. Losses in the frequency selective surface are introduced by printing the periodic pattern through resistive inks and hence avoiding the typical soldering of a large number of lumped resistors. The effect of the surface resistance of the FSS and dielectric substrate characteristics on the input impedance of the absorber is discussed by means of a circuital model. It is shown that the optimum value of surface resistance is affected both by substrate parameters (thickness and permittivity) and by FSS element shape. The equivalent circuit model is then used to introduce the working principles of the narrowband and the wideband absorbing structure and to derive the best-suited element for wideband absorption. Finally, the experimental validation of the presented structures is presented.

## I. INTRODUCTION

Recent developments in the synthesis of novel metamaterials, exhibiting anomalous electromagnetic properties, have increased interest in planar frequency selective surfaces, because of their great versatility and simple manufacturing procedures. These surfaces can be employed in front of a grounded dielectric slab to synthesize high-impedance surfaces [1]-[3]. Several attempts to employ these complex materials for improving the performance of classical electromagnetic absorbers have been made in last years [4]-[17]. This technology allows approaching the physical limitations about the thickness bandwidth ratio [18], [19]. Thin narrow band absorbers employing metallic FSS loaded with a large number of lumped resistors have been repeatedly proposed [6]-[9]. Loaded HIS surfaces can be also employed to synthesize thin and wideband absorbers [10]-[16]. The use of lumped resistors for introducing ohmic losses in a metallic surface leads to complex and expensive structures because of the cost of high frequency resistors and complexity of the manufacturing. An attractive alternative to this approach consists in printing the frequency selective surface through resistive inks with a proper surface resistance [12]-[14]. Some design rules for designing resistive loaded HIS absorbers are presented in [17] where they are referred to as circuit analog absorbers. However, the theory in the book is specific to the wideband design and experimental verifications are not shown. A specific theory of narrowband structures and a useful methodology for choosing the right surface resistance of the FSS is not available. A practical realization of resistive patterns employed in thin absorbers can be found in [13] and in [14]. Nevertheless, in these papers, a non optimal design is obtained due to absence of designing principles.

Here we present an analysis of the absorbing structure by a simple equivalent circuit that allows a detailed explanation of the key parameters for the design of the absorber. We show that the optimum surface resistance of the ink composing the FSS depends on the unit cell shape, on the distance of the FSS from the ground plane and on the substrate permittivity. Finally, the experimental verification of the absorbing structure is presented.

The paper is organized as follows: in the next Section, the structure under analysis is described by means of an equivalent circuit approach; in Sect. III the relation between the FSS surface resistance and the lumped resistance of the equivalent circuit is clarified. Sect. IV shows an example of thin absorber for narrow-band absorption. Some considerations about the bandwidth of the device, keeping constant the dielectric thickness, are presented in order to determine the best-suited FSS element. In Sect. V the theory of the wideband configuration is presented together with a practical design example. Finally in Sect. V the experimental results of both the narrowband and the wideband configuration are shown.

## II. PROPOSED CONFIGURATION AND CIRCUIT MODEL

The absorbing panel consists of a conventional high-impedance surface comprising lossy frequency selective surfaces over a thin grounded dielectric slab (see Fig. 1). The FSS array, made up of capacitive cells, behaves as a capacitor in the low frequency region but its impedance becomes inductive after the first resonance.

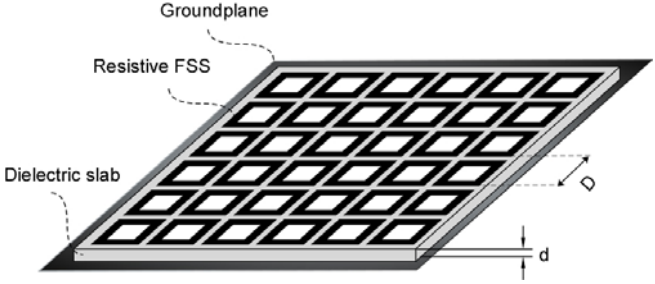

**Fig. 1** – Three-dimensional sketch of the analyzed configuration.

The impedance of a lossy FSS can be represented through a series RLC circuit:

$$Z_{FSS} = R - j\left(\frac{1-\omega^2 LC}{\omega C}\right) \quad (1)$$

The shape of FSS influences the values of L and C parameters. The surface impedance of the absorbing structure $Z_R$ is equal to the parallel connection between the FSS impedance $Z_{FSS}$ and the surface impedance of the grounded dielectric slab $Z_d$:

$$Z_R = \frac{Z_d Z_{FSS}}{(Z_d + Z_{FSS})} \quad (2)$$

The thin grounded dielectric slab behaves as an inductor. Its impedance can be computed analitycally as follows:

$$Z_d = jZ_m^{TE,TM}\tan(\beta d) \quad (3)$$

where $Z_m^{TE} = (\omega\mu_r\mu_0)/\beta$; $Z_m^{TM} = \beta/(\omega\varepsilon_r\varepsilon_0)$ are the characteristic impedances of the slab for TE and TM polarization, $\beta = \sqrt{k^2 - k_t^2}$ is the propagation constant along the normal unit of the slab and $k_t = k_0\sin(\vartheta)$ is the transverse wavenumber with $\theta$ representing the incidence angle of the incoming wave with respect to the normal.

After some analytical manipulations, the real and the immaginary part of the input impedance $Z_R$ can be derived:

$$\text{Re}\{Z_R\} = \frac{RZ_d^2}{\left[\frac{1-\omega^2 LC}{\omega C} - Z_d\right]^2 + R^2} \quad (4)$$

$$\text{Im}\{Z_R\} = \frac{[-\omega L Z_d]\cdot\left[\frac{1-\omega^2 LC}{\omega C} - Z_d\right] + \left[\frac{1}{\omega C}\cdot Z_d\right]\cdot\left[\frac{1-\omega^2 LC}{\omega C} - Z_d\right] + R^2 Z_d}{\left[\frac{1-\omega^2 LC}{\omega C} - Z_d\right]^2 + R^2} \quad (5)$$

When the inductive impedance of the substrate $Z_d$ and the immaginary part of the FSS impedance $\left((1-\omega^2 LC)/\omega C\right)$ assume the same value, the parallel circuit resonates and the impedance $Z_R$ becomes purely real (for thin substrates) and equal to:

$$\text{Re}\{Z_R\} = \frac{\left(Z_m^{TE,TM}\right)^2\tan^2(\beta d)}{R} \quad (6)$$

In order to obtain absorption of the incoming signal, the impedance expressed in (6) should match the free space impedance $\zeta_0$:

$$R_{opt} = \frac{\left(Z_m^{TE,TM}\right)^2\tan^2(\beta d)}{\zeta_0^{TE,TM}} \quad (7)$$

The relation in (7) highlights the dependence of the optimal FSS resistance on thickness and permittivity of the substrate. From (7), it is evident that the thicker is the substrate (given by $d$), the higher is the optimal FSS surface resistance.

### III. RELATION BETWEEN LUMPED CIRCUIT RESISTANCE $R$ AND FSS SURFACE RESISTANCE $R_S$

In order to derive the optimal surface resistance from the equivalent circuit, a relation between the lumped resistance $R$ of the equivalent circuit and the surface resistance of the FSS $R_S$ has to be defined. In a strict sense, the optimal surface resistance of the lossy layer would be exactly equal to $R_{opt}$ if the ink coating were uniform. If the layer is not uniform, as in the case of an FSS, the surface resistance depends not only on the parameters defined in (7), namely thickness and dielectric constant of the substrate, but more remarkably, also on the unit cell shape. In particular, the scattering surface area per unit cell is larger in the uniform sheet model than the physical scatterer [20]. As a good estimate, the addition of losses is accounted by a series load resistance of value:

$$R \approx R_s \frac{S}{A} \quad (8)$$

where $S=D^2$, $D$ is the cell periodicity and $A$ is the surface area of the lossy element within a single unit cell. The relation (8) implies that the smaller is the scattering area, the smaller is the surface resistance that generates a certain fixed lumped resistance (same amount of losses). This formula holds very well for patch type FSSs but, when resonant shapes such as rings or crosses are considered, the surface area $A$ is represented only by the surface area of the element along the direction of the current (parallel to the incoming electric field). In a cross FSS, for instance, such area is the area of one of the two crossed dipoles composing the element. In the case of a ring shaped FSS, the relation (8) slightly underestimates the lumped resistance $R$ since the current in the ring is split in two segments parallel to the electric field. Indeed, considering a cross and a ring element with the same surface area, it can be easily realized that the current density on the two narrow strips of the ring is higher than the current density on the cross element. Nevertheless, a rigorous calculation of the lumped resistance can be performed by imposing the power dissipated in the equivalent transmission line equal to the power lost in the unit cell of the lossy element. For a freestanding lossy FSS, the power dissipated in the resistor of the equivalent transmission line reads:

$$P_L = \frac{1}{2}\left|\frac{V^+}{\varsigma_0}\right|^2 [1-\Gamma]\left|\frac{\varsigma_0}{R+\varsigma_0+jX}\right|^2 R \quad (9)$$

where $V^+$ represents the incoming voltage directly proportional to the incident electric field, $\zeta_0$ is the free space impedance, $\Gamma$ is the reflection coefficient computed in correspondence of the position of the FSS, $R$ represents the lumped resistance of the of the FSS and $X$ is the reactance due to the inductance and the capacitance of the specific element. The power lost in the unit cell of the real structure is computed as follow:

$$P_D = R_S \iint_{unitcell} |J(x,y)|^2 \, dxdy \qquad (10)$$

Alternatively, the exact value of lumped resistance can be obtained after a full wave simulation of a freestanding FSS by computing the real part of the FSS impedance [21]. Such parameter is constant up to frequencies where the unit cell is reasonably homogeneous and it corresponds to the value of the resistance in the lumped circuit.

The approximate lumped model here presented allows to acquire a valuable insight into the physical principles of the structure. In order to validate the effectiveness of the formulas, in Fig. 2 the real and the imaginary part of the input impedance of the absorber, $Z_R$, are shown. An impinging wave is absorbed by the resistive HIS when the input impedance matches the free space impedance. The purpose of the graph is to show that, for a given surface resistance of the FSS (e.g. $R_S$=50 Ω/sq), an increase of the substrate thickness (keeping constant the operating frequency and therefore the wavelength) leads to an increase of the real part of $Z_R$ according to (4). The considered FSS shape is a square patch with a ratio $P$ between the side length $L$ and the cell periodicity $D$ equal to 10/16. The lumped resistance of the approximate RLC circuit is computed according to (8) and the L, C values have been obtained by employing the inversion method proposed in [21] (C=23 fF, L=1.49 nH @ D=10 mm). In order to obtain an increase of the thickness with respect to the wavelength, the same resonance frequency is maintained by changing the unit cell periodicity.

This variation leads to a rescaling of the L and C lumped parameters [21] but, as predicted by (4) and (6), their values do not influence the maximum value of $\text{Re}\{Z_R\}$. The surface resistance of the patches is equal to 50 Ω/sq in all cases.

## IV. NARROW BAND CONFIGURATION

The presented approach can be employed to synthesize an ultra-thin narrow band absorber. The best-suited FSS shape for achieving the largest operating band with a fixed substrate (thickness and permittivity) is the patch element with a narrow gap between adjacent metalized squares [22].

In Fig. 3 the reflection coefficients of the HIS absorber obtained with different cell shapes are shown. The lossy frequency selective surface is printed on a grounded air slab with a thickness of 1 mm corresponding to $\lambda/30$ at the operating frequency. It can be noticed that the optimal surface resistance is higher for cells with a high percentage of printed surface within the unit cell. Indeed, according to (8), a cell with a small ratio between the surface area of the lossy element and the unit cell area *(A/S)* generates the same value of equivalent lumped resistance $R$ with a smaller surface resistance.

The use of substrates with high dielectric permittivity generates angular stabilization of the operating frequency but also a strong bandwidth reduction. The employment of vias results a good solution to enlarge and stabilize the bandwidth of the absorber for oblique TM polarization [23].

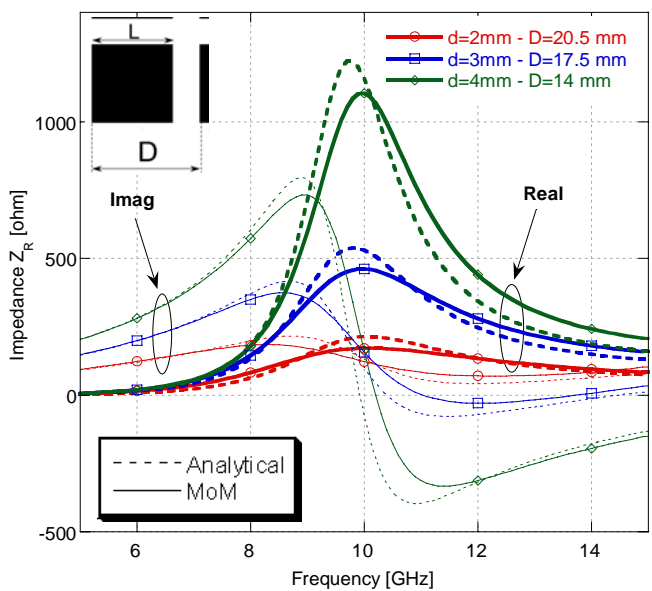

**Fig. 2** – Real and imaginary part of the impedance $Z_R$ with different thicknesses of the substrate. The FSS shape is a patch with a ratio P=L/D of 10/16. The surface resistance is always equal to 50 Ω/sq.

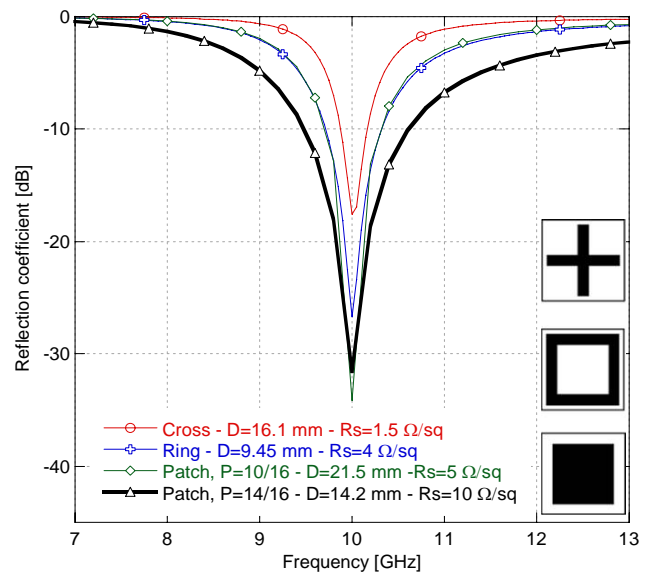

**Fig. 3** – Reflection coefficient of the narrow-band HIS absorber with different cell shapes. Substrate parameters: thickness=1 mm, $\varepsilon_r$=1.

## V. WIDEBAND CONFIGURATION

As highlighted through the circuital approach, in correspondence of the resonance frequency generated by the parallel between the capacitive impedance of the FSS and the inductive impedance of the grounded dielectric slab, the value of the real part of the input impedance increases. This allows to synthesize a good absorber in correspondence of the HIS resonances. By choosing an FSS shape that resonates well before the grating lobes, as for instance crosses or rings, a second parallel resonance between the FSS and the grounded substrate can be generated. Such resonance is obtained when the inductive impedance of the FSS assumes the same value of the capacitive impedance of the substrate. As is well known, a grounded dielectric substrate behaves as a capacitance when its thickness becomes thicker than a quarter wavelength. It is therefore necessary employing a slightly thicker substrate to obtain a capacitive behavior in the frequency range of interest. The choice of a rather thick substrate guarantees a sufficiently high value of the real part of $Z_R$ also between the two parallel resonances. In addition, at the center of the operating band, the substrate thickness equals $\lambda/4$ acting as a high-impedance wall while the FSS impedance turns from capacitive to inductive behavior showing a purely real impedance. These circumstances allow the structure working as a conventional Salisbury screen [24], [25] at the center of the operating band [17]. The working principles of the structure can be also graphically explained considering a test example. Let us consider an FSS composed by a ring array, with a periodicity $D$ equal to 11 mm and a surface resistance $R_s$ of 70 Ω/sq, printed on a air grounded dielectric substrate with a thickness of 5 mm (see the three-dimensional sketch of the analyzed absorbing structure reported in Fig. 1)

In Fig. 4 the impedance of the grounded dielectric substrate and the impedance of the FSS, computed by retrieving full wave data of reflection coefficient, are reported. The two resonances previously mentioned and the Salisbury screen zone are highlighted in the figure. By increasing the value of $R_s$, the real part of $Z_R$ is lowered in correspondence of the resonances and increased between them. An optimal value of $R_s$ to obtain a wide frequency range where the real part of $Z_R$ is around 400 Ω corresponds to 70 Ω/sq. The second maximum of $\text{Re}\{Z_R\}$ falls before the second resonance of the HIS since its value is dominated by the square of the substrate impedance (see relation (6)) that has a deep minimum below the second resonance. In Fig. 5 the reflection coefficient of the absorber obtained by a periodic MoM code, by the equivalent circuit approach and by Ansoft HFSS v.10 is reported. In the circuital approach the lumped resistance is computed with a moderate correction to the relation (8) in agreement to what argued in the section III. The absorbing structure allows obtaining remarkable performance (-15 dB in the band from 7 GHz to 20 GHz) with an overall thickness of 5 mm only. This performance cannot be accomplished by lightweight configurations employing optimized Jaumann screen [26] or by other commercially available non-magnetic multilayer structures (see for instance [27]) with a thickness lower than 9-10 mm.

Despite the intrinsic periodicity of the structure, its dimension can be reduced down to a 4 by 4 array preserving almost the same absorption performances, with respect to a PEC plate of the same dimensions.

The absorber here presented does not redirect the energy in other directions as in other RAM designs [28] but it dissipates the incoming power by realizing the matching condition over a wide frequency range. The energy would be reflected in other directions only if the FSS period were larger than one wavelength (as it happens in the reference [13]). In the present design the redirection of the energy toward the grating lobes starts after 27 GHz.

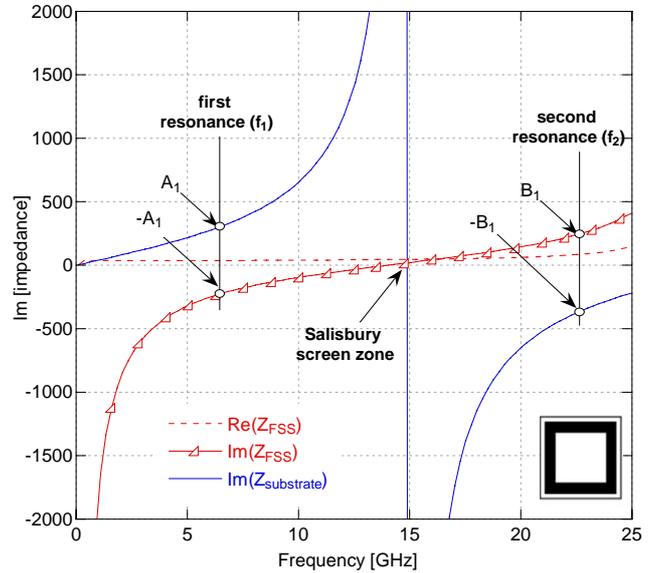

**Fig. 4** – Impedance of a 5mm grounded air slab and impedance of a ring shaped FSS with periodicity D=11mm. The resonances of the HIS are highlighted.

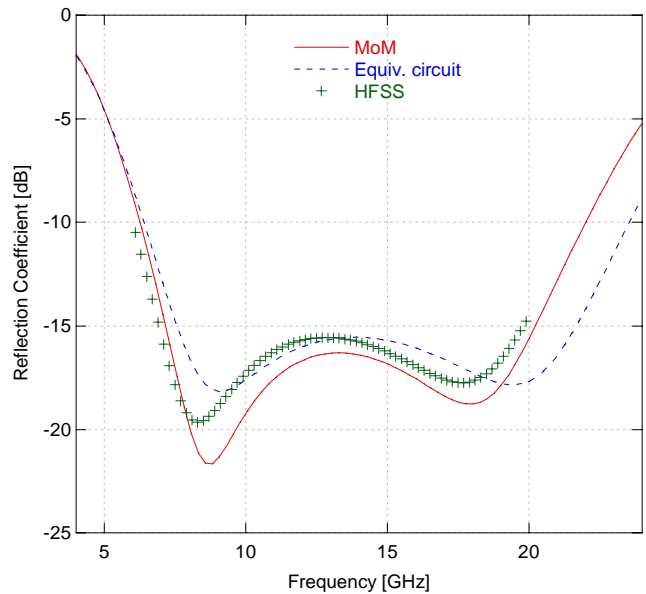

**Fig. 5** – Reflection coefficient of the analyzed structure: ring shaped FSS over a grounded air slab of 5mm.

For the sake of completeness we report in Fig. 6 the behavior of the absorber for oblique incidence.

The absorption of the structure is acceptable up to 30 degrees but it starts to deteriorate after 45 degrees. However, it is worth to underline that the aim of the present paper is to present and discuss the absorption mechanisms of this kind of structure, and not to optimize an absorber for oblique incidence angles. Methods for improving the performance of the absorber for oblique incidence consists in employing an additional upper dielectric as in [11] or by optimizing the design through an iterative technique.

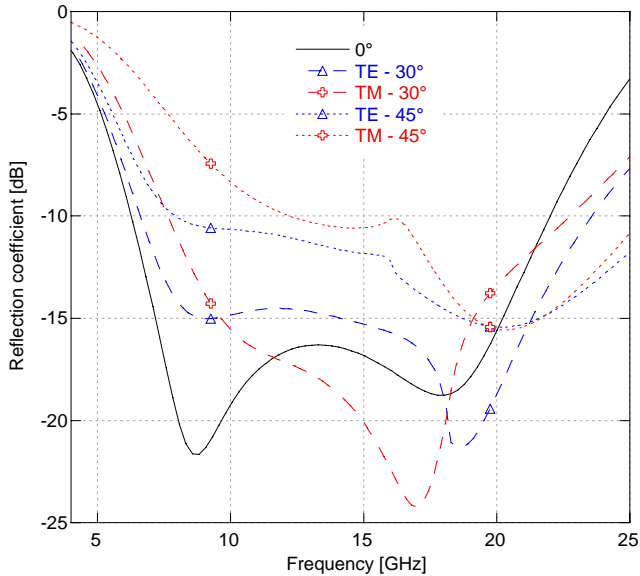

**Fig. 6 -** Reflection coefficient for oblique incidence angles of the analyzed absorber based on the ring FSS over a 5 mm air slab.

## VI. EXPERIMENTAL VERIFICATION

The feasibility of the thin absorbing structures is demonstrated by two experimental prototypes. The resistive patterns representing the lossy frequency selective surfaces have been manufactured by the silk printing technique through a photo etched frame. A picture of the narrowband sample is shown in Fig. 7 and the measured reflection coefficient, compared with simulated result, is plotted in Fig. 8.
In this example, the FSS consists of a patch array with a periodicity of 10 mm and a ratio *P* between the side length *L* of the element and the cell periodicity *D* equal to 12/16. As shown in section IV, the use of the patch type FSS guarantees the largest possible bandwidth [22].

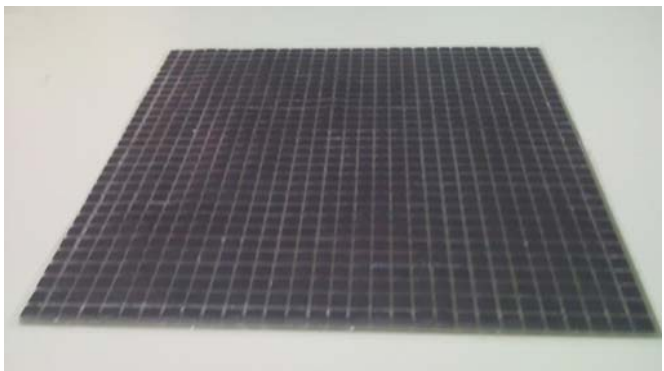

**Fig. 7** – A picture of a narrow band absorber. The overall thickness is equal to 2.4 mm.

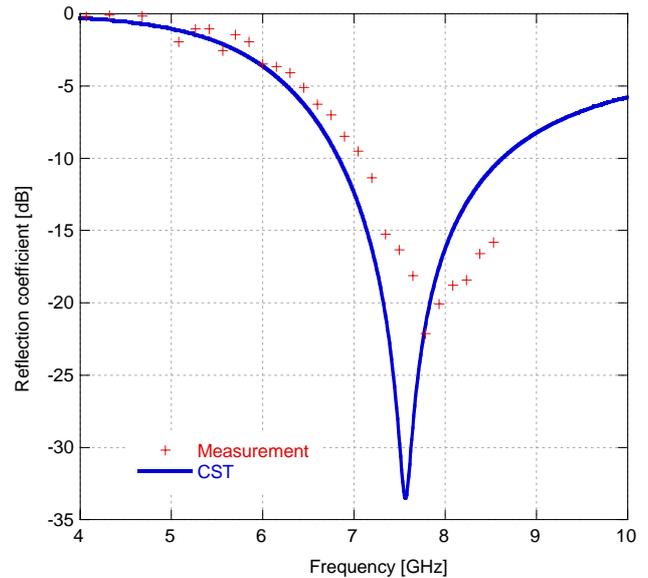

**Fig. 8** – Measured reflection coefficient of the sample shown in Fig. 7 compared to the computed result. Measured data are reported up to 8.6 GHz because of limitation of the available network analyzer.

The surface resistance of the square patches is equal to 35 Ω/sq and the substrate is the commercial FR4 with permittivity equal to $\varepsilon_r$=4.5-j0.088. The overall thickness of the absorber is 2.4 mm that corresponds to $\lambda_g/8$ at the resonant frequency. The bandwidth is comparable to the one achievable by a conventional Salisbury configuration.

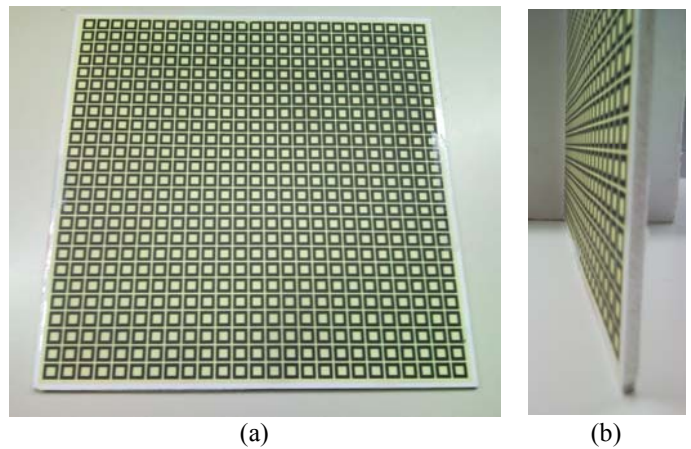

(a)                                                              (b)

**Fig. 9** - A picture of a wideband absorber. Overall thickness equal to 5 mm. (a) front view, (b) side view.

The prototype of the wideband configuration is similar to the one previously analyzed in the section V. Both a Kapton substrate and a slim FR4 substrate of 0.15 mm have been used to pattern the resistive FSS. The use of FR4 substrate requires a slight modification of the ring periodicity (12 mm). The air spacer has been created by employing a Rohacell substrate with a relative permittivity of 1.05-j0.017. A picture of the wideband sample is shown in Fig. 9. The manufactured structure results in a very lightweight configuration, indeed a 30 cm × 30 cm sample weighs 10 g compared with the 450 g of a commercial magnetically loaded absorber (Eccosorb [29]) with the same dimensions.

In Fig. 10 we present the measured reflection coefficient of two samples realized by different paints (one declared for 70 Ω/sq end the other one for 10 Ω/sq) compared with the computed results. It results that the first prototype, referred as prototype *A*, is characterized by a surface resistance of roughly 130 Ω/sq and the second configuration (prototype *B*) is characterized by a surface resistance of 20 Ω/sq.

The results are not the optimal ones presented in the previous simulations but they are presented in order to verify the design methodology. The paints employed in the manufacturing process need to be mixed for obtaining the right surface resistance value, but this represents a practical issue that goes beyond the scope of this paper.

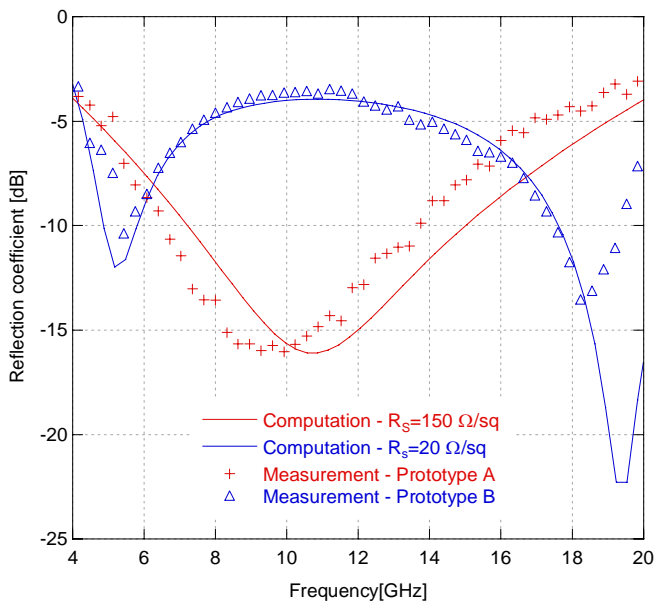

**Fig. 10** - Measured Reflection coefficient of the sample in Fig. 9 compared with the computed result.

## VII. CONCLUSION

In this paper we have provided a detailed explanation of thin absorbers based on high-impedance surfaces. The absorber consists of a lossy frequency selective surface placed above a thin grounded dielectric slab. The absorbing structure can be employed for synthesizing both thin narrowband and wideband absorbers that strongly outperform the conventional Salisbury and Jaumann configurations. In the present work, the lossy FSS has been synthesized by printing a resistive pattern trough resistive inks. The use of resistive patterns provides a dramatically simpler and lightweight structure with respect to the design including a large number of lumped resistors. By means of a lumped equivalent circuit, we have determined simple rules for choosing the optimal surface resistance of the FSS. Indeed, this latter represents the key parameter of the structure. We have showed that its optimal value depends on the FSS shape and on the substrate parameters. The working principles of thin narrowband and wideband absorbers have been addressed by the same model. Computational and experimental results have been presented to verify the two analyzed configurations.

## VIII. APPENDIX – SURFACE RESISTANCE

The surface resistance is defined to represent a slab of lossy material, with thickness $\Delta$ and conductivity $\sigma$. A resistive sheet whose properties are entirely specified by the single measurable quantity $Z_s$ (Ω/square) can be analyzed as an infinitesimal sheet by a MoM code. However, various definitions of this parameter are available in literature. When the thickness of the sheet is much larger than the wavelength the well known expression for computing the surface impedance is valid:

$$Z_s = (1+j) \cdot \frac{1}{(\delta \sigma)} \quad \text{(A1)}$$

where $\sigma$ is the conductivity of the material and $\delta$ represents the skin depth. The expression of the skin depth of a material reads:

$$\delta = \sqrt{2/(\omega \mu \sigma)} \quad \text{(A2)}$$

where $\omega$ is the operating radian frequency and $\mu$ is the permeability of the material. When the conductivity of the material is lowered, the skin depth of the material becomes rather larger than the sheet thickness. In this case a boundary condition for thin slabs has to be employed. In [25] the surface impedance of a thin homogeneous slab (of thickness $\Delta$) is defined as follows:

$$Z_s = \frac{1}{\omega \cdot \Delta \cdot j\left(\varepsilon_0 \varepsilon_r - j\sigma/\omega\right)} \quad \text{(A3)}$$

under two restrictions: $\varepsilon_r \mu_r \gg 1$, $k_0 \varepsilon_r \mu_r \Delta^2/2 \ll 1$. From (A3) it results that, if $\varepsilon_r'' \gg \varepsilon_r'$ condition is verified, $Z_s$ is always positive and its imaginary part is negligible.

Another formulation for the surface impedance boundary condition was introduced in 1903 by Levi-Cività and Picciati [30]. By rewriting the relation in a form similar to (A3) we obtain:

$$Z_s = \frac{1}{\omega \cdot \Delta \cdot j\left(\varepsilon_0 (\varepsilon_r - 1) - j\sigma/\omega\right)}. \quad \text{(A4)}$$

valid on the assumption that $k_0 \Delta \ll 1$ $\varepsilon_r \mu_r \gg 1$. The relations (A3) and (A4) differ only for a correction on the permittivity of the material that does not cause appreciable variations. It is worth underlining that, when the imaginary part of the dielectric permittivity is much larger than the real part, the relation (A3) and (A4) are well approximated by the static definition of the surface impedance:

$$Z_s = 1/(\sigma \Delta) \quad \text{(A5)}$$

In the analyzed frequency range, this approximation is valid if

$$\sigma \gg \omega \varepsilon_0 \varepsilon_r \approx 0.5 \cdot \varepsilon_r\big|_{10GHz} \quad \text{(A6)}$$

Considering a resistive FSS with a thickness of 25 μm (typical for commercial inks), this condition leads to a superior limit of roughly 1000 Ω/sq (if $\varepsilon_r$=1) for replacing the condition (A5) to the more general (A3) and (A4). Beyond this limit, the value of the surface impedance is not anymore purely real.

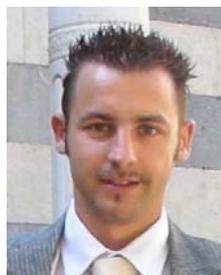

**Filippo Costa** (S '07) was born in Pisa, Italy, on October 31, 1980. He received the M.Sc. degree in telecommunication engineering and the Ph.D. degree in applied electromagnetism in electrical and biomedical engineering, electronics, smart sensors, nano-technologies from the University of Pisa, Pisa, Italy, in 2006 and 2010, respectively. He is currently a Post Doc researcher at the University of Pisa. From March to August 2009, he was a Visiting Researcher at the Department of Radio Science and Engineering, Helsinki University of Technology, TKK, Finland. His research is focused on the analysis and modelling of artificial impedance surfaces with emphasis on their application in electromagnetic absorbers and low-profile antennas.

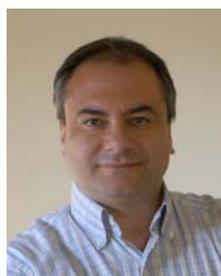

**Agostino Monorchio** (S'89–M'96–SM'04) received the Laurea degree in electronics engineering and the Ph.D. degree in methods and technologies for environmental monitoring from the University of Pisa, Pisa, Italy, in 1991 and 1994, respectively. During 1995, he joined the Radio Astronomy Group, Arcetri Astrophysical Observatory, Florence, Italy, as a Postdoctoral Research Fellow, in the area of antennas and microwave systems. He has been collaborating with the Electromagnetic Communication Laboratory, Pennsylvania State University (Penn State), University Park, and he is an Affiliate of the Computational Electromagnetics and Antennas Research Laboratory. He has been a Visiting Scientist at the University of Granada, Spain, and at the Communication University of China in Beijing. He is currently an Associate Professor in the School of Engineering, University of Pisa, and Adjunct Professor at the Italian Naval Academy of Livorno. He is also an Adjunct Professor in the Department of Electrical Engineering, Penn State. He is on the Teaching Board of the Ph.D. course in "Remote Sensing" and on the council of the Ph.D. School of Engineering "Leonardo da Vinci" at the University of Pisa.

His research interests include the development of novel numerical and asymptotic methods in applied electromagnetics, both in frequency and time domains, with applications to the design of antennas, microwave systems and RCS calculation, the analysis and design of frequency-selective surfaces and novel materials, and the definition of electromagnetic scattering models from complex objects and random surfaces for remote sensing applications. He has been a reviewer for many scientific journals and he has been supervising various research projects related to applied electromagnetics commissioned and supported by national companies and public institutions.

Dr. Monorchio has served as Associate Editor of the IEEE ANTENNAS AND WIRELESS PROPAGATION LETTERS. He received a Summa Foundation Fellowship and a NATO Senior Fellowship.

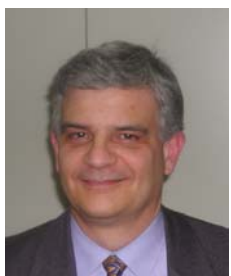

**Giuliano Manara** was born in Florence, Italy, on October 30, 1954. He received the Laurea (Doctor) degree in electronics engineering (summa cum laude) from the University of Florence, Italy, in 1979. Currently, he is a Professor at the College of Engineering of the University of Pisa, Italy. Since 2000, he has been serving as the President of the Bachelor and the Master Programs in Telecommunication Engineering at the same University. Since 1980, he has been collaborating with the Department of Electrical Engineering of the Ohio State University, Columbus, Ohio, where, in the summer and fall of 1987, he was involved in research at the ElectroScience Laboratory.

His research interests have centered mainly on the asymptotic solution of radiation and scattering problems to improve and extend the uniform geometrical theory of diffraction. In this framework, he has analyzed electromagnetic wave scattering from material bodies, with emphasis on the scattering from both isotropic and anisotropic impedance wedges. He has also been engaged in research on numerical, analytical and hybrid techniques (both in frequency and time domain), scattering from rough surfaces, frequency selective surfaces (FSS), and electromagnetic compatibility. More recently, his research has also been focused on the design of microwave antennas with application to broadband wireless networks, and on the development and testing of new microwave materials (metamaterials).

Prof. Manara was elected an IEEE Fellow in 2004 for "contributions to the uniform geometrical theory of diffraction and its applications." Since 2002, he has been serving as a member of the IEEE Italy Section Executive Committee. In May 2004, Prof. Manara was the Chairman of the Local Organizing Committee for the International Symposium on Electromagnetic Theory of Commission B of the International Union of Radio Science (URSI). He also served as a Convenor for several URSI Commission B international conferences, and URSI General Assemblies. In August 2008, he has been elected Vice-Chair of the International Commission B of URSI.